\documentclass[aps,prb,twocolumn,floats,showpacs]{revtex4}

\usepackage{epsfig}

\usepackage{amsmath}

\usepackage{amssymb}

\usepackage{bm}

\def\be{\begin{equation}}
\def\ee{\end{equation}}
\def\bea{\begin{eqnarray}}
\def\eea{\end{eqnarray}}

\begin{document}

\title{Canted Magnetization Texture in Ferromagnetic Tunnel Junctions}

\author{Igor Kuzmenko and Vladimir Fal'ko}
\affiliation{Department of Physics, Lancaster University,
        Lancaster, LA1 4YB, United Kingdom}
\date{\today}

\begin{abstract}
We study the formation of inhomogeneous magnetization texture in
the vicinity of a tunnel junction between two ferromagnetic wires
nominally in the antiparallel configuration and its influence on
the magnetoresistance of such a device. The texture, dependent on
magnetization rigidity and crystalline anisotropy energy in the
ferromagnet, appears upon an increase of ferromagnetic inter-wire
coupling above a critical value and it varies with an external
magnetic field.
\end{abstract}


\maketitle


Spin polarized transport in multilayer
ferromagnet-metal-ferromagnet system and magnetic tunnel junction
is a subject of intense theoretical and experimental studies.
\cite{Printz98,Brataas06} The majority of such studies addresses
the investigation of the tunneling magneto-resistance
\cite{Julliere75,Moodera00,Usai01} (MR) and giant MR
\cite{Gerber97,Fert88,Grunberg89,Pratt91-Lee92,GMR-APL06} effects,
which consist of a switch from lower to higher conductivity when
polarization of leads in a MR device change from an antiparallel
to parallel configuration. In tunnel junctions, the MR effect is
the result of a difference between rates of tunneling of electrons
majority and minority bands on the opposite sides of a junctions,
and it's the strongest when magnetization switching by an external
magnetic field changes from an ideally antiparallel magnetization
state in the two wires to a parallel one.
\cite{Moodera95,Zutic04,Mathon-prb01,Li98} Any deviations of the
magnetization near the interface (where tunneling characteristics
of the device are formed) from perfectly parallel/antiparallel
orientation would reduce the size of the effect.

\begin{figure}[htb]
\centering
\includegraphics[width=75mm,height=52mm,angle=0]{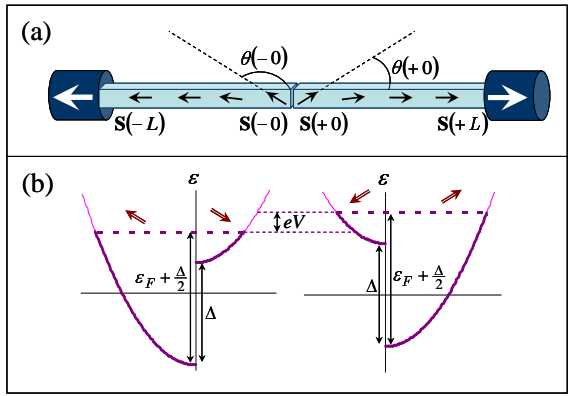}
 \caption{{\bf{(a)}} Ferromagnetic wires. Polarizations
          of the wire ends $z=0$ and $z=\pm{L}$ are
          ${\bf{S}}(\pm0)$ and ${\bf{S}}(\pm{L})$,
          respectively. The angle between ${\bf{S}}(\pm0)$
          and the axis $z$ is $\theta(\pm0)$.
          {\bf{(b)}} Band alignment of the majority and minority
          bands for electrons in the vicinity of the left and
          right-hand side of a biased ferromagnetic junction
          in the antiparallel configuration. Direction of spin
          quantization on either side of the junction is determined
          by the orientation of ${\bf{S}}(\pm0)$ of magnetization
          near the interface.}
 \label{Fig-SpinChain}
\end{figure}

In this paper we investigate a possibility of formation of
inhomogeneous magnetization texture in the vicinity of a highly
transparent tunnel junction caused by ferromagnetic coupling of
magnetic moments on the opposite sides carried by tunneling
electrons. We find that a canted magnetization state can form if
such ferromagnetic tunneling coupling, $t'$, exceeds some critical
value $t_{\rm{c}}$ determined by the interplay between crystalline
anisotropy and magnetization rigidity in the ferromagnet. This
means that a tunnel junction with $t'<t_{\rm{c}}$ can be viewed as
an atomically sharp magnetic domain wall, whereas the increase of
the junction transparency above $t_{\rm{c}}$ gradually transforms
it into a broad texture typical for bulk ferromagnetic material.
For $t'>t_{\rm{c}}$, we study the evolution of the texture upon
application of an external magnetic field and construct a
parametric diagram for distinct magnetization regimes. As an
example, we consider a device consisting of a tunnel junction
between two easy-axis ferromagnetic wires magnetically biased at
the ends as illustrated in Fig.\ref{Fig-SpinChain}. We show that
when the magnetic field exceeds some critical value, $B_{\rm{c}}$,
the barrier between the states with polarization
parallel/antiparallel to the field disappears. As a result, both
wires will be polarized parallel to the magnetic field and the
domain wall will be pushed away toward the end of the wire. When a
magnetic field is swept back (to $B<B_{\bf{c}}$) and its sign
changes, the domain wall returns back to the tunnel junction. The
resulting hysteresis in the magnetization state of the device
leads to a hysteresis loop in its MR, which we analyze taking into
account the formation of the texture near the tunnel junction with
$t'>t_{\rm{c}}$.

A quasi-1D magnetization texture, ${\bf{S}}(z)$, near the tunnel
junction between two wires of length $L$, which changes slowly on
the scale of the lattice constant $a$ can be described using the
energy functional,
\begin{widetext}
\begin{eqnarray}
  E &=&
  \frac{J}{2}
  \Bigg(
       \int\limits_{-L}^{-0}+
       \int\limits_{+0}^{L}
  \Bigg)dz
  \Big(\partial_z{\bf{S}}(z)\Big)^2-
  \frac{w^2}{a^3}
  \int\limits_{-L}^{L}dz
  \bigg[
       \xi_1\Big(S^z(z)\Big)^2+
       \xi_2\Big(S^y(z)\Big)^2
  \bigg]+
  \nonumber \\ && +
  \frac{g_dw^3}{2a^3}
  \int\limits_{-L}^{L}\frac{dz dz'}{w^2}
  V(z-z')
  \bigg\{
       S^x(z)S^x(z')+
       S^y(z)S^y(z')-
       2S^z(z)S^z(z')
  \bigg\}-
  \frac{\mu B w^2}{a^3}
  \int\limits_{-L}^{L}{dz}
  S^{z}(z)+
  E'.
  \label{E-tot}
\end{eqnarray}
\end{widetext}
Here $J\sim{t}w^2/a$ is the magnetization rigidity (where $t$ is
the exchange coupling between the neighboring atoms, $w$ is the
wire thickness). The crystalline anisotropy parameters are
$\xi_1>\xi_2>0$, whereas $g_d=\gamma^2/a^3$ parameterizes
dipole-dipole interaction of magnetic moments, with
\begin{eqnarray*}
  V(z) &=&
  \frac{1}{2w}
  \int{d^2{\boldsymbol{\rho}}}
  {d^2{\boldsymbol{\rho}}'}
  \frac{2z^2-(x-x')^2-(y-y')^2}
       {\big((x-x')^2+(y-y')^2+z^2\big)^{5/2}},
\end{eqnarray*}
where the integration is carried out over the cross-section of the
wire, $d^2{\boldsymbol{\rho}}=dxdy$,
$d^2{\boldsymbol{\rho}}'=dx'dy'$. For $|z|>w$, $V(z)$ decreases as
$w^3/z^3$, and for $|z|\ll{w}$, $V(z)\propto\ln(w/|z|)$. For a
smooth magnetization texture varying at a length scale longer than
the wire thickness, we approximate the non-local dipole-dipole
interaction as $V(z) \approx V_0w\delta(z)$,
$V_0=\int\frac{dz}{w}V(z).$ Finally, $\mu$ is the magnetic moment
per atom, and ${\bf{B}}=B{\bf{e}}_z$ is an external magnetic
field.

The inter-wire coupling energy, $E'$, includes the exchange
interaction due to the penetration of the polarized electron wave
function through the barrier, from one ferromagnetic metal into
another. For the ferromagnetic sign of the inter-wire coupling,
$$
 E'=
 -\frac{t'w^2}{a^2}
 \Big(
     {\bf{S}}(+0)
     \cdot
     {\bf{S}}(-0)
 \Big).
$$

In the following, we will focus on the magnetization texture
formed near the tunnel junction between two ferromagnetic metals
with antiparallel polarization ${\bf{S}}(\pm{L})=\pm{\bf{e}}_z$
fixed by magnetic reservoirs at the distant ends of the wires.
Without coupling between wires ($t'=0$), this would determine
homogeneous magnetization ${\bf{S}}(z>0)={\bf{e}}_z$ and
${\bf{S}}(z<0)=-{\bf{e}}_z$. The exchange interaction between the
wires may give rise to the formation of canted magnetization
texture in the vicinity of the junction, with boundary values of
spin ${\bf{S}}(\pm0)$ determined by the interplay between the
magnetization rigidity, crystalline anisotropy, Zeeman energy and
the exchange inter-wire interaction. In the case when the easy
magnetization direction is along $z$ axis, we parameterize
\begin{equation}
  {\bf{S}}(z)=
  {\bf{e}}_z\cos(\theta(z))+
  {\bf{e}}_y\sin(\theta(z)),
  \label{S-polar}
\end{equation}
with $\theta(-L)=\pi$ and $\theta(L)=0$. Then, the total energy of
the interacting wires takes the form
\begin{eqnarray}
  E &=&
  \frac{J}{2}
  \bigg(
  \int\limits_{-L}^{-0}+
  \int\limits_{+0}^{L}
  \bigg)dz
  \bigg\{
       (\theta'(z))^2+
       \alpha^2\sin^2(\theta(z))+
  \nonumber \\ && +
       2\alpha^2\lambda_B
       \Big[
           1-\cos(\theta(z))
       \Big]
  \bigg\}-
  t'\cos(\theta_0),
  \label{E-cont-angle}
\end{eqnarray}
where $\theta'(z)=d\theta(z)/dz$, $\theta_0=\theta(-0)-\theta(+0)$.
Relevant parameters present in Eq.(\ref{E-cont-angle}) are
\begin{eqnarray*}
  \alpha=
  \sqrt{\frac{w^2}{a^3}
        \frac{2(\xi_1-\xi_2)+3g_dV_0}{J}},
  \ \ \
  \lambda_B=
  \frac{2\mu B w^2}{J a^3 \alpha^2}
  \equiv
  \frac{B}{B_{\rm{c}}},
\end{eqnarray*}
where
$$ B_{\rm{c}}=\frac{J a^3 \alpha^2}{2 \mu w^2}. $$
The meaning of the parameter $B_{\rm{c}}$ is the following. When
$B=B_{\rm{c}}$, the energy barrier between the states of
noninteracting wires ($t'=0$) with polarizations
parallel/antiparallel to the external magnetic field disappears
and the polarization antiparallel to the magnetic field becomes
absolutely unstable, leading to a jump of the magnetic domain wall
from the junction toward the magnetically biased end of the wires.
Below we assume that $B_{\rm{c}}$ is  much less than the field
required to switch the polarization of the left/right hand side
leads. The parameters $\alpha$ and $\lambda_B$ characterize a
typical domain wall width in an infinite wire. For example,
$\alpha^{-1}$ is the domain wall width for $B=0$ for a long wire
with $L\alpha\gg1$. An external magnetic field
${\bf{B}}=B{\bf{e}}_z$ makes the domain wall asymmetric. It
compresses the domain wall width on the side where magnetization
is aligned with ${\bf{B}}$ to $\alpha_{+}^{-1}$,
$\alpha_{+}=\alpha\sqrt{1+\lambda_B}$, and it stretches the domain
wall width on the side where ${\bf{S}}$ is antiparallel to
${\bf{B}}$ to $\alpha_{-}^{-1}$,
$\alpha_{-}=\alpha\sqrt{1-\lambda_B}$.

To minimize the energy in Eq. (\ref{E-cont-angle}), we employ the
following procedure. First, we fix the boundary values
$\theta(\pm0)$ and find the optimal form of $\theta(z)$ for given
$\theta(\pm0)$. Then we minimize $E(\theta(+0),\theta(-0))$ versus
the canting angles $\theta(\pm0)$. The first step of such a
procedure requires solving the optimum equation
\begin{eqnarray}
  \theta''(z)=
  \frac{\alpha^2}{2}
  \Big[
      \sin(2\theta(z))+
      2\lambda_B\sin(\theta(z))
  \Big].
  \label{Eq-mot-B}
\end{eqnarray}
The latter shows that that $\theta(z>0)$ [$\pi-\theta(z<0)$] takes
its maximal value $\theta(+0)$ [$\pi-\theta(-0)$] at $z=0$, and
that it decreases as $\exp(-\alpha_{+}z)$ [$\exp(\alpha_{-}z)$]
for $|z|\gg\alpha^{-1}$. Also the differential equation
(\ref{Eq-mot-B}) has the first integral
$v_{\pm}(\theta)=-\theta'(z)$,
\begin{eqnarray}
  v_{\pm}(\theta)=
  \alpha
  \sqrt{(1\pm\lambda_B)^2-
        (\cos(\theta)+
         \lambda_B)^2},
  \label{v-theta-B}
\end{eqnarray}
where the sign $\pm$ corresponds to positive/negative $z$.
Substituting $v_{\pm}$ into energy in Eq. (\ref{E-cont-angle}) and
changing the integration variable from $z$ to $\theta$ we arrive
at
\begin{eqnarray}
  E &=&
  J
  \Bigg(
       \int\limits_{0}^{\theta(+0)}v_{+}(\theta)d\theta+
       \int\limits_{\theta(-0)}^{\pi}v_{-}(\theta)d\theta
  \Bigg)-
  \nonumber \\ && -
  t'\cos(\theta_0),
  \label{E-v-Bpar}
\end{eqnarray}
which represents the explicit dependence of energy
$E(\theta(+0),\theta(-0))$ on the canting angles on each side of
the junction. Minimizing the energy with respect to these angles,
we identify possible regimes for the magnetization texture.

First, we consider the magnetization texture in the absence of an
external magnetic field, $B=0$. In this case the texture is
symmetric, $\theta(-z)=\pi-\theta(z)$, so that
$v_{+}(\theta)=v_{-}(\theta)=\alpha\sin(\theta)$ and
$$
 E=2J\alpha(1-\cos(\theta(+0)))+t'\cos(2\theta(+0)).
$$
The latter result indicates the existence of a critical value
$t_{\rm{c}}=J\alpha/2$ of the coupling constant $t'$, such that
for $t'<t_{\rm{c}}$ the energy reaches its minimum when
$\theta(+0)=\pi-\theta(-0)=0$ and magnetization is homogeneous in
each of two wires, whereas for $t'>t_{\rm c}$ the energy minimum
corresponds to the magnetization texture in the
vicinity of the tunnel junction, with
\begin{eqnarray*}
 \theta(+0)=\pi-\theta(-0)=
 {\rm{arccos}}(t_{\rm c}/t').
\end{eqnarray*}

In the presence of external magnetic field, the magnetization
texture becomes asymmetric. In this case, we determine canting
angles $\theta(\pm0)$ numerically from the set of two equations,
\begin{eqnarray}
  \partial_{\theta(\pm0)}E=
  J v_{\pm}(\theta(\pm0))-t'\sin(\theta(-0)-\theta(+0)) &=& 0.
  \nonumber \\
  \label{Eq-rectangular}
\end{eqnarray}

\begin{figure}[htb]
\centering
\includegraphics[width=55mm,height=49mm,angle=0]{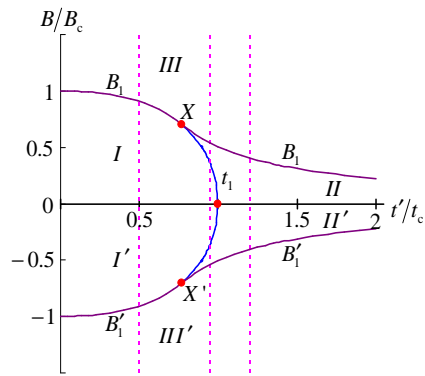}
\caption{Parametric diagram $t'$-$B$. There are four
         lines, $B_1$, $B'_1$, $t_1$, and the axis $B=0$
         separating phases one from another (see the text for
         details). The lines $B_1$ and $t_1$ ($B'_1=-B_1$
         and $t_1$) touch one another at the point $X$ ($X'$).
         The boundary values $\theta(\pm0)$ as a function of
         the external magnetic field are calculated for
         $t'=0.5t_{\rm{c}}$, $0.95t_{\rm{c}}$ and $1.2t_{\rm{c}}$
         (dashed lines).}
 \label{Fig-Phase-Diagram}
\end{figure}

The results of numerical analysis of Eqs.(\ref{Eq-rectangular})
are gathered in the parametric diagram in Fig.
\ref{Fig-Phase-Diagram}, where we indicate the six parametric
intervals separated by lines $B_1$, $B'_1$, $t_1$, and the axis
$B=0$ corresponding to different regimes of magnetization texture.
Below, we describe the evolution of the magnetization state of the
junction upon sweeping the magnetic field from $-B_{\rm{c}}$ to
$+B_{\rm{c}}$ (in comparison to an inverse sweep from
$+B_{\rm{c}}$ to $-B_{\rm{c}}$). When the exchange coupling
constant $t'$ and the magnetic field $B$ are not strong enough
(parametric interval $I$ ($I'$) in Fig.\ref{Fig-Phase-Diagram}),
the energy functional has two minima. The first of them
corresponds to $\theta(+0)=\pi-\theta(-0)=0$ and the magnetization
is homogeneous in each of two wires. The second corresponds to the
state with both wires polarized parallel to the external magnetic
field and the domain wall pushed away towards the end $z=-L$
($z=L$) of the wire. As a result, the polarizations of the wires
in the tunneling area are antiparallel/parallel one to another
depending on whether we increase the magnetic field from $B=0$ or
decrease the field from $B_1$ (decrease the field from $B=0$ or
increase the field from $B'_1$). When one decreases the magnetic
field from $B_1$ (increases from $B'_1$) so that the field changes
its sign, the state with homogeneous polarization of both wires
and the domain wall placed at the end $z=-L$ ($z=L$) of the wire
becomes unstable, and the domain wall is pushed towards the
tunneling area $z=0$.

Changing $t'$ and/or $B$ and crossing the line $t_1$ results in
continuous variation of the canting angles from
$\theta(+0)=\pi-\theta(-0)=0$ on the left from the line $t_1$ to
finite values on the right from the line $t_1$ (parametric
interval $II$ ($II'$) in Fig.\ref{Fig-Phase-Diagram}) and
formation of the magnetization texture near the tunnel junction.
The values of the canting angles are found from equations
(\ref{Eq-rectangular}). The second state corresponds to the
situation when both the wires are polarized parallel to the
external magnetic field and the domain wall is pushed away towards
the end $z=-L$ ($z=L$) of the wire. At the line $B_1$ ($B'_1$) the
barrier between metastatic and ground states disappears and there
is just one minimum of the energy functional corresponding to the
state with both wires polarized parallel to the external magnetic
field and the domain wall pushed away towards the end $z=-L$
($z=L$) of the wire. As a result, the polarization in the vicinity
of the tunnel junction is homogeneous, i.e.,
$\theta(-0)=\theta(+0)=0$ ($\theta(-0)=\theta(+0)=\pi$).

\begin{figure}[htb]
  \centering
  \includegraphics[width=60mm,height=100mm,angle=0]{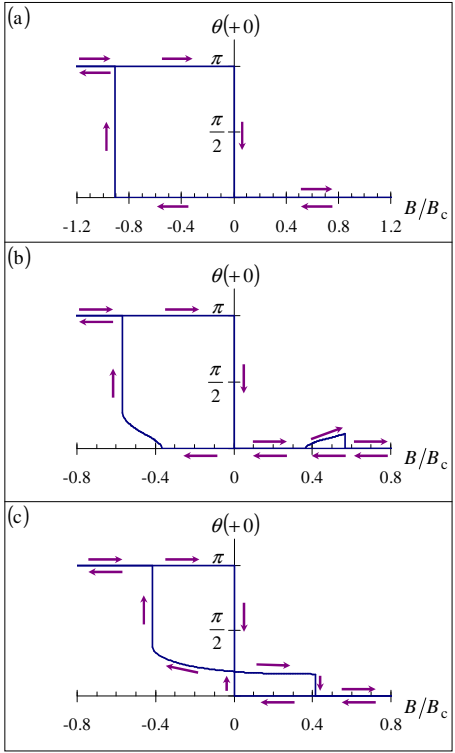}
  \caption{Canting angle $\theta(+0)$ as functions of $\lambda_B$
           for $t'=0.5t_{\rm{c}}$ (panel (a)),
           $t'=0.95t_{\rm{c}}$ (panel (b)), and
           $t'=1.2t_{\rm{c}}$ (panel (c)).}
 \label{Fig-th-R}
\end{figure}

Typical dependencies of $\theta(+0)$ on a magnetic field $B$
are illustrated in Fig. \ref{Fig-th-R} for $t'=0.5t_{\rm{c}}$,
$t'=0.95t_{\rm{c}}$, and $t'=1.2t_{\rm{c}}$, respectively. The
dependencies of $\theta(-0)$ on $B$ can be obtained from
Fig.~\ref{Fig-th-R} by the transformation
$$ \theta(-0,B)=\pi-\theta(+0,-B). $$

The formation of canted magnetization in the vicinity of a tunnel
junction with high transparency would affect the magnetoresistance
characteristic of such a junction. The latter can be modelled
using the tunnel Hamiltonian approach. When the domain wall width
$\alpha^{-1}$ is sufficiently larger then the elastic mean free
path, the tunneling Hamiltonian can be written as
\begin{eqnarray*}
  H = H_0+H_T,
  & \ \ \ &
  H_0=
  \sum_{\nu=\pm}\sum_{{\bf{k}},\sigma}
  \epsilon_{\sigma}({\bf{k}})
  c_{\nu{\bf{k}}\sigma}^{\dag}
  c_{\nu{\bf{k}}\sigma},
\end{eqnarray*}
where
$\epsilon_{\sigma}({\bf{k}})=\epsilon({\bf{k}})-\sigma\Delta$.
Here $H_0$ is the Hamiltonian of the isolated ferromagnetic wires,
$H_T$ is the tunneling Hamiltonian, $c_{\nu{\bf{k}}\sigma}$ and
$c^{\dag}_{\nu{\bf{k}}\sigma}$ are annihilation and creation
operators of electron propagating in the left ($\nu=L$) or right
($\nu=R$) wire with wave vector ${\bf{k}}$ and spin parallel
($\sigma=1/2$ or $\uparrow$) or antiparallel ($\sigma=-1/2$ or
$\downarrow$) to the wire polarization ${\bf{S}}(\pm0)$
(\ref{S-polar}). In the following we will assume that the Fermi
momenta for the majority and minority bands are sufficiently
larger than $\alpha_{\pm}$ and therefore treat conduction band
electrons as three dimensional.
\begin{eqnarray*}
  H_T=
  \sum_{\sigma,\sigma'}
  \sum_{{\bf{k}},{\bf{k}}'}
  \bigg[
       t_{\sigma'\sigma}({\bf{k}}',{\bf{k}})
       c^{\dag}_{R{\bf{k}}'\sigma'}
       c_{L{\bf{k}}\sigma}+
       {\rm h.c.}
  \bigg],
\end{eqnarray*}
where the tunneling matrix elements
$t_{\sigma'\sigma}({\bf{k}}',{\bf{k}})$ describe the transfer of
an electron with wave vector ${\bf{k}}$ and spin state $\sigma$
from the left wire to the state with ${\bf{k}}'$ and $\sigma'$ in
the right wire and the quantization axes for the conduction band
electrons are directed along the magnetization vectors
${\bf{S}}(\pm0)$ (\ref{S-polar}) which are not necessarily
collinear so that the transitions between the majority/minority
bands of one wire and majority/minority bands of the other are
possible even for the nominally antiparallel configuration of
${\bf{S}}(\pm{L})$. We consider the model in which electrons
tunnel from one wire to another without spin flipping. In this
case the spin dependence of
$t_{\sigma'\sigma}({\bf{k}}',{\bf{k}})$ is determined by a single
parameter, the angle $\theta_0=\theta(-0)-\theta(+0)$ between the
vectors ${\bf{S}}(\pm0)$,
\begin{eqnarray*}
  t_{\sigma\sigma'}({\bf{k}}',{\bf{k}}) &=&
  \tilde{t}
  \bigg|
       \frac{2\pi^2\hbar^2v^z({\bf{k}})v^z({\bf{k}}')}{L^2}
  \bigg|^{1/2}
  \times \nonumber \\ &\times&
  \Big[
      \cos(\theta_0/2)\delta_{\sigma\sigma'}+
      i\sin(\theta_0/2)\tau^x_{\sigma\sigma'}
  \Big],
\end{eqnarray*}
where $\tau^x$ is the Pauli matrix, and $v^z({\bf{k}})$ is a
component of electron velocity
${\bf{v}}({\bf{k}})=\nabla_{\bf{k}}\epsilon_{\sigma}({\bf{k}})/\hbar$
perpendicular to the interface.

When vectors ${\bf{S}}(+0)$ and ${\bf{S}}(-0)$ are antiparallel
(parametric intervals $I$ and $I'$ in
Fig.\ref{Fig-Phase-Diagram}), electrons can tunnel only from
the majority band of one wire to the minority band of the
other,\cite{Falko02} so that conductance of such a junction is
\begin{eqnarray*}
  &&
  G_{\uparrow\downarrow}=
  \frac{2\pi e^2|\tilde{t}|^2}{\hbar}
  ~
  \bigg(
       \frac{2\pi\hbar}{L}
  \bigg)^2
  N_{\uparrow}v^{z}_{\uparrow}
  N_{\downarrow}v^{z}_{\downarrow},
\end{eqnarray*}
where $N_{\uparrow}$ and $N_{\downarrow}$ are the densities of
states in the majority and minority bands at the Fermi level, and
$v^z_{\uparrow}$/$v^z_{\downarrow}$ are the average value of
$|v^{z}({\bf{k}})|$ over the majority/minority Fermi surface,
\begin{eqnarray*}
  v^z_{\sigma} &=&
  \frac{1}{N_{\sigma}} \
  \sum_{{\bf{k}}}
  \Big|
      v^z({\bf{k}})
  \Big|
  \delta
  \Big(
      \epsilon-
      \epsilon({\bf{k}})
  \Big).
\end{eqnarray*}
When the ends $z=\pm0$ of the wires have parallel magnetization,
i.e., $\theta(\pm0)=0$ or $\pi$, the conductance is
\begin{eqnarray*}
  &&
  G_{\uparrow\uparrow}=
  \frac{2\pi e^2|\tilde{t}|^2}{\hbar}
  ~
  \bigg(
       \frac{2\pi\hbar}{L}
  \bigg)^2
  \frac{(N_{\uparrow}v^{z}_{\uparrow})^2+
       (N_{\downarrow}v^{z}_{\downarrow})^2}
      {2},
\end{eqnarray*}
which determines the magneto-resistance (MR) ratio
\begin{eqnarray*}
  \delta_0=
  \frac{G_{\uparrow\uparrow}-G_{\uparrow\downarrow}}
       {G_{\uparrow\uparrow}}.
\end{eqnarray*}

\begin{figure}[htb]
\centering
\includegraphics[width=85mm,height=92mm,angle=0]{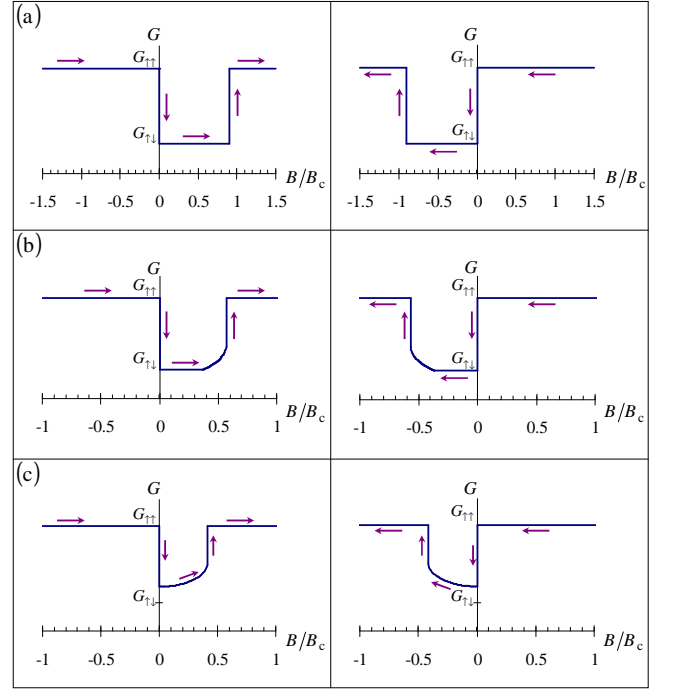}
 \caption{Magnetoresistance for a positive (left) and negative
 (right) magnetic field sweep, showing the effect of canted
 magnetization texture on the form of hysteresis in the magnetic
 tunnel junction, for: {\bf(a)} $t'=0.5t_{\rm{c}}$,
 {\bf(b)} $t'=0.95t_{\rm{c}}$, and {\bf(c)} $t'=1.2t_{\rm{c}}$.}
 \label{Fig-G-hyst}
\end{figure}

The formation of a canted magnetization texture in the vicinity of
a junction with angle $\theta_0$ between magnetization directions
of the opposite sides of the tunnel barrier produces conductance
$G(\theta_0)$ and the reduced observable MR ratio, $\delta$,
\begin{eqnarray}
  &&
  G(\theta_0)=
  G_{\uparrow\uparrow}
  \cos^2(\theta_0/2)+
  G_{\uparrow\downarrow}
  \sin^2(\theta_0/2),
  \nonumber
  \\
  &&
  \delta=
  \frac{G_{\uparrow\uparrow}-G(\theta_0)}
       {G_{\uparrow\uparrow}}=
  \delta_{0}\sin^2(\theta_0/2).
  \label{delta}
\end{eqnarray}

The results of calculations for the conductance in MR devices with
various junction transparencies (and positive and negative
magnetic field sweep in the magnetoresistance) are gathered in
Fig.\ref{Fig-G-hyst}. The conductance dip at small magnetic fields
indicates the regime when a ferromagnetic domain wall in the
device in Fig.\ref{Fig-SpinChain} is pinned to the tunnel
junction. The detailed structure of the hysteresis loop depends on
whether the magnetization texture is formed near the junction or
not which of course depends on the value of the ferromagnetic
inter-wire coupling. For small inter-wire coupling, in
Fig.\ref{Fig-G-hyst}(a), the jump between parallel and
antiparallel polarizations in the vicinity of the tunnel junction
give rise to jumps in the conductance between
$G_{\uparrow\uparrow}$ and flat conductance minimum equal to
$G_{\uparrow\downarrow}$. With increasing the inter-wire coupling
towards the critical value $t_{\rm{c}}$ (determined by the
interplay between crystalline anisotropy and magnetization
rigidity in ferromagnet), an interval of magnetic fields appears
where the conductance (\ref{delta}) increases continuously, due to
the formation of the canted magnetization texture and its change
by a magnetic field, Fig.\ref{Fig-G-hyst}(b). For inter-wire
coupling above a critical value $t_{\rm{c}}$,
Fig.\ref{Fig-G-hyst}(c), the minimum of the conductance exceeds
$G_{\uparrow\downarrow}$ even in the absence of a magnetic field,
indicating the formation of a broad texture, which also suggests a
reduction of MR ratio in a high-transparency junction.

The authors thank to G. Bauer and E. McCann for useful discussions.
This project has been funded by EU STREP Dynamax and ESFCRP
``Spin current''.

\end{document}